\documentclass[referee,a4paper]{raa}            
\usepackage[margin=2.65cm]{geometry}
\usepackage{graphicx,times}             
\usepackage{natbib}
\usepackage{amssymb,amsmath}
\bibpunct{(}{)}{;}{a}{}{,}
\usepackage[pagebackref=true]{hyperref}
\usepackage{amsmath}
\usepackage{rotating,tabularx}

\newcommand{\kms}{{\rm km\,s^{-1}}}
\newcommand{\lya}{Ly$\alpha$}

\newcommand{\NHI}{N_{\rm HI}}
\newcommand{\logNHI}{\log N_{\rm HI}}

\newcommand{\HI}{\ion{H}{1}}

\begin{document}
\title{Constraining the Temperature-Density Relation of the Inter-Galactic Medium from Analytically Modeling Lyman-alpha Forest Absorbers}
   \volnopage{Vol.0 (20xx) No.0, 000--000}      
   \setcounter{page}{1}          

   \author{Li Yang 
      \inst{1,2,3}
   \and Zheng Zheng 
      \inst{3}
   \and T.-S. Kim
      \inst{4}
   }

   \institute{Shanghai Astronomical Observatory, Chinese Academy of Sciences,
80 Nandan Road, Shanghai 200030, People's Republic of China; {\it liyang@shao.ac.cn}\\
        \and
             School of Astronomy and Space Sciences, University of Chinese Academy of Sciences,
No.19A Yuquan Road, Beijing 100049, People’s Republic of China \\
        \and
             Department of Physics and Astronomy, University of Utah,
115 S 1400 E, Salt Lake City, UT 84112, USA; {\it zhengzheng@astro.utah.edu}\\
        \and 
            Department of Astronomy, University of Wisconsin,
475 North Charter Street, Madison, WI 53706, USA \\
\vs\no
   {\small Received 20xx month day; accepted 20xx month day}}
%
%
\abstract{ 
The absorption by neutral hydrogen in the intergalactic medium (IGM) produces the \lya\ forest in the spectra of quasars. The \lya\ forest absorbers have a broad distribution of neutral hydrogen column density $\NHI$ and Doppler $b$ parameter. The narrowest \lya\ absorption lines (of lowest $b$) with neutral hydrogen column density above $\sim 10^{13}{\rm cm^{-2}}$ are dominated by thermal broadening, which can be used to constrain the thermal state of the IGM. Here we constrain the temperature-density relation $T=T_0(\rho/\bar{\rho})^{\gamma-1}$ of the IGM at $1.6<z<3.6$ by using $\NHI$ and $b$ parameters measured from 24 high-resolution and high-signal-to-noise quasar spectra and by employing an analytic model to model the $\NHI$-dependent low-$b$ cutoff in the $b$ distribution. 
In each $\NHI$ bin, the $b$ cutoff is estimated using two methods, one non-parametric method from computing the cumulative $b$ distribution and a parametric method from fitting the full $b$ distribution.
We find that the IGM temperature $T_0$ at the mean gas density $\bar{\rho}$ shows a peak of $\sim 1.5\times 10^4$K at $z\sim $2.7--2.9. At redshift higher than this, the index $\gamma$ approximately remains constant, and it starts to increase toward lower redshifts. The evolution in both parameters is in good agreement with constraints from completely different approaches, which signals that \ion{He}{2} reionization completes around $z\sim 3$.
}
   \authorrunning{Li Yang, Zheng Zheng \& T.-S. Kim }  
   \titlerunning{Temperature-Density Relation of the IGM from Lyman-alpha Forest Absorbers }  

   \maketitle
%
%
\section{Introduction}
\label{section::Introduction}
\setcounter{footnote}{0}

The Lyman-$\alpha$ (\lya) forest, namely the ensemble of absorption lines blueward of the \lya\ emission in the spectra of the quasar, is caused by the absorption of intervening neutral hydrogen in the intergalactic medium (IGM) \citep[e.g.,][]{Cen94,Bi97,Rauch98}. As the largest reservoir of baryons, the evolution of the IGM is affected by several processes, such as adiabatic cooling caused by cosmic expansion, heating by ionizing photons from galaxies and quasars, and heating from gravitational collapse. 
The \lya\ forest encodes the thermal state of the IGM \citep{Gunn1965,Lynds1971,Hui1997,Schaye1999,Schaye2000} and therefore it has become the premier probe of the thermal and ionization history of the IGM.

The thermal state of the IGM is usually characterized by the temperature-density relation, parameterized as $T=T_0\Delta^{\gamma-1}$ \citep{Hui1997}. Here, $\Delta\equiv \rho/\bar{\rho}$ is the ratio of the gas density to its cosmic mean, the normalization $T_0$ corresponds to the temperature of the gas at the mean density, and $\gamma$ denotes the slope of the relation. Measuring $T_0$ and $\gamma$ as a function of redshift would allow the reconstruction of the thermal and ionization history of the IGM. In particular, the \ion{H}{1} reionization and \ion{He}{2} reionization leave distinct features in the evolution of $T_0$ and $\gamma$, and the properties of the \lya\ forest contain their imprint \citep[e.g.,][]{Miralda1994,Theuns2002,Hui2003,Worseck2011,Puchwein2015,Upton2016,Worseck2016,Gaikwad2019,Worseck2019,Upton2020}. Various statistical properties of the \lya\ forest have been applied to measure $T_0$ and $\gamma$, such as the \lya\ forest flux power spectrum and the probability distribution of flux (\citealt{Theuns2000,McDonald2001,Zaldarriaga2001,Bolton2008,Viel2009,Calura2012,Lee2015,Rorai2017,Walther2018,Boera2019,Khaire2019,Walther2019}; see \citealt{Gaikwad2021} for a summary). 

There is also a method of measuring the IGM thermal state based on Voigt profile decomposition of the \lya\ forest \citep[e.g.,][]{Schaye1999,Schaye2000,Ricotti2000,Bryan2000,McDonald2001,Rudie2012,Bolton2014,Rorai2018,Hiss2018}. In this approach, the \lya\ absorption spectrum is treated as a superposition of multiple discrete Voigt profiles, with each line described by three parameters: redshift $z$, Doppler parameter $b$, and neutral hydrogen column density $\NHI$. By studying the statistical properties of these parameters, i.e., the $b$--$\NHI$ distribution at a given redshift, one can recover the thermal information encoded in the absorption profiles. The underlying principle of this approach is that the narrow absorption lines (with low $b$) are dominated by thermal broadening, determined by the thermal state of the IGM.

Such an approach involves the determination of the cutoff in the $b$ distribution as a function of $\NHI$. The commonly-used method follows an iterative procedure introduced by \citet{Schaye1999}: fit the observed $b$--$\NHI$ distribution with a power law; discard the data points 1$\sigma$ above the fit; iterate the procedure until convergence in the fit. With the small number of \HI\ lines around the $b$ cutoff and contamination by noise and metals, different line lists can lead to different results on the $T_0$ and $\gamma$ constraints  \citep[e.g.,][]{Rudie2012,Hiss2018}.

In this paper, we employ the low-$b$ cutoff profile approach to constrain the IGM thermal state at $z\sim 3$. We circumvent the above problem of determining the profile of the $b$ cutoff by proposing two different methods. The first one is a non-parametric method, which measures the lower 10th percentile in the $b$ distribution from the cumulative $b$ distribution in each $\NHI$ bin. The other one is a parametric method, which infers the lower 10th percentile in the $b$ distribution from a parametric fit to the full $b$ distribution in each $\NHI$ bin. With the determined $b$ cutoff profiles, to derive the constraints on $T_0$ and $\gamma$, we apply a physically motivated and reasonably calibrated analytic model describing the $b$ cutoff, which avoids using intensive simulations. Unlike previous work \citep[e.g.,][]{Schaye1999,Rudie2012,Hiss2018}, where the low-$b$ cutoff profile is obtained by iteratively removing data points based on power-law fits, our methods make use of a well-defined $\NHI$-dependent low-$b$ cutoff threshold, i.e., the lower 10th percentile in the $b$ distribution. Adopting such a quantitative cutoff threshold allows a direct comparison to the analytic model of \citet{Garzilli2015} with the same cutoff threshold, providing a simple and efficient way of constraining the thermal state of the IGM. In this paper, we presnt the methods and apply them for the first time to observed \lya\ forest line measurements for $T_0$ and $\gamma$ constraints.

In Section~\ref{sec:data}, we describe the data used in this work, which is a list of \lya\ absorption lines with Voigt profile measurements from 24 observed high-resolution and high-signal-to-noise (high-S/N) quasar spectra by \citet{Kim2021}. In Section~\ref{sec:constraints}, we present the overall distribution of $\NHI$ and $b$. Then we present the two methods of determining the $b$ cutoff profile and obtain the constraints on $T_0$ and $\gamma$ in the redshift range of $1.6<z<3.6$. Finally, we summarize our results in Section~\ref{sec:summary}. In the Appendix~\ref{sec:appendix}, we list the constraints in Table~\ref{table_T0gamma}. 

%
%
\section{Data Samples and Reduction Methods}
\label{sec:data}

In this work, we analyze the fitted line parameters of the \lya\ forest by \citet{Kim2021}: the absorber redshift $z$, the (logarithmic) neutral hydrogen column density $\log \NHI=\log[\NHI/({\rm cm^{-2}})]$, and the Doppler parameter $b\, {\rm (\kms)}$. 
This list is based on Voigt profile fitting analysis for the twenty-four high-resolution and high-S/N quasar spectra, taken from HIRES (HIgh-Resolution Echelle Spectrometer; \citealt{Vogt1994,Vogt2002}) on Keck I and UVES (UV-Visible Echelle Spectrograph; \citealt{Dekker2000}) on the VLT (Very Large Telescope). The resolution is about 6.7 $\kms$. The list of quasars and the details of the fitting analysis can be found in \citet{Kim2021}.

There are two sets of the fitted parameters, one using only the \lya\ absorption (the \lya-only fit) and the other using all the available Lyman series lines (the Lyman series fit). 
The 24 UVES/HIRES quasar spectra provide 5615 (6638) \HI\ lines at $1.6<z<3.6$ for the Lyman series (\lya-only) fit.
The Lyman series fit can derive more reliable line parameters for saturated \lya\ lines at $\logNHI \gtrsim 14.5$. Note that even including all the available high-order Lyman series lines does not vouch for the completely resolved profile structure of heavily saturated lines at $\log\NHI \gtrsim$ 17--18, if severe line blending and intervening Lyman limit systems leave no clean high-order Lyman series lines. The $\NHI$ detection limit is about $\log\NHI\sim 12.5$. At $\log\NHI \in [13.5,16.0]$, where the incompleteness is negligible, the 24 UVES/HIRES quasar spectra provide 1810 (2058) \HI\ lines at $1.6<z<3.6$ for the Lyman series (\lya-only) fit. In our analysis, we use both sets of parameters.

The data set from 24 HIRES/UVES quasar spectra in \citet{Kim2021} is unique in combining three aspects of the line fitting analysis: high-S/N ($>$45 per pixel) to reduce the possibility of misidentifying metal lines as \ion{H}{I} absorption lines, without Damped \lya\ Absorbers (DLAs) in the spectra to avoid cutting down the available wavelength region significantly and the difficulty in spectra normalization and removal of metals blended with \ion{H}{I}, and Voigt profile fitting both from \lya\ only and from available Lyman series to more effectively deblend saturated lines. As a comparison, in studying the IGM thermal state, \citet{Hiss2018} use 75 HIRES/UVES spectra at $2.0<z<3.4$ with low S/N ($>$15 per pixel) containing DLAs, with parameters derived from \lya-only fit; \citet{Gaikwad2021} use 103 HIRES spectra at $2.0<z<4.0$ with low S/N ($>$5 per pixel) without DLAs or sub-DLAs, also with parameters derived from \lya-only fit; \citet{Rudie2012} use 15 HIRES spectra with high S/N ($>$50 per pixel) containing DLAs, with parameters from fitting both \lya\ and Ly$\beta$, but only covering $2<z<2.8$. The line measurements in \citet{Kim2021} from the self-consistent, uniform in-depth Voigt profile fitting analysis with reduced systematics are well suited to our application of testing new methods of studying the IGM around $z\sim 3$.

\begin{figure*}
\centering
\includegraphics[width=.49\textwidth]{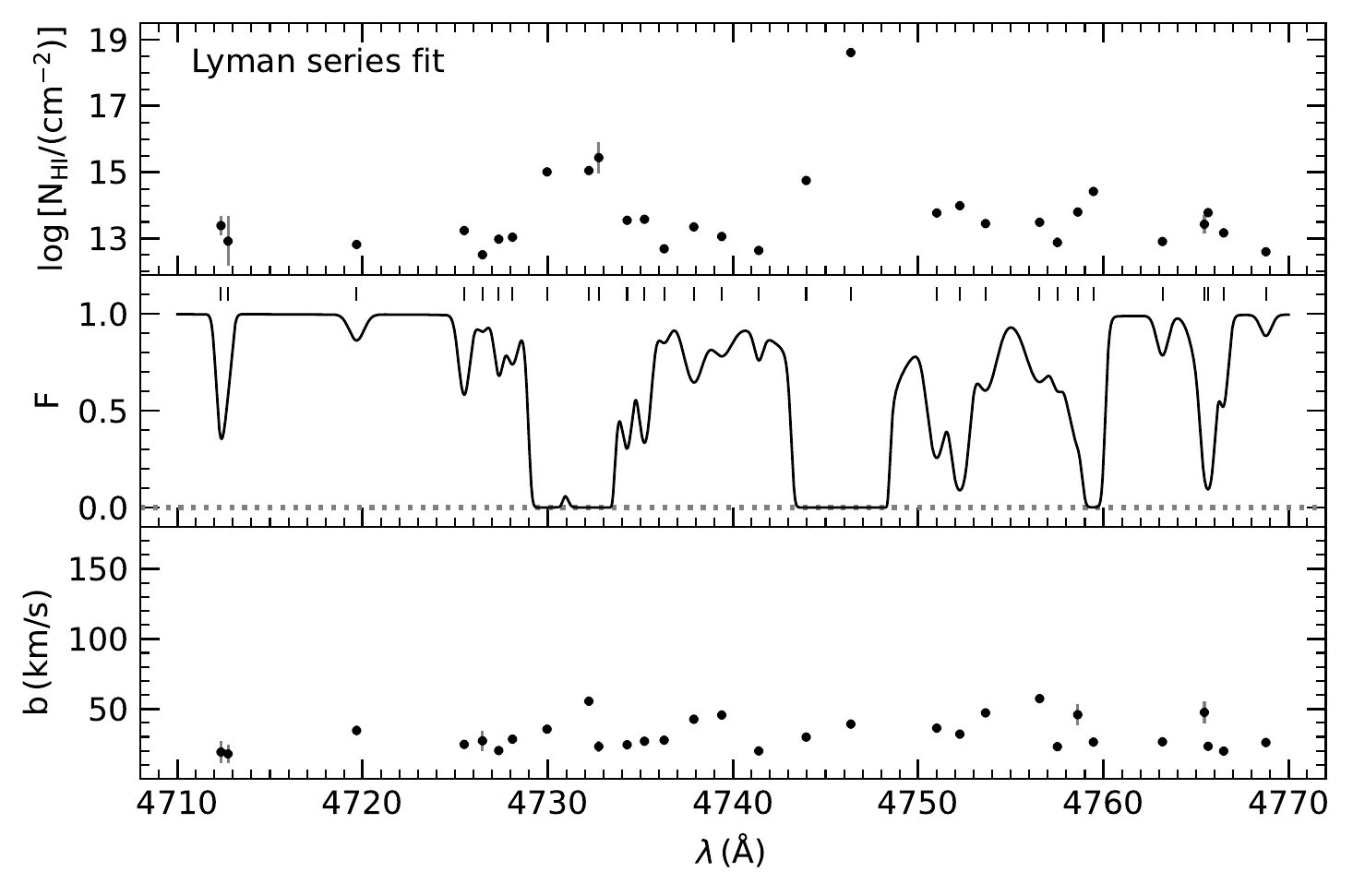}
\includegraphics[width=.49\textwidth]{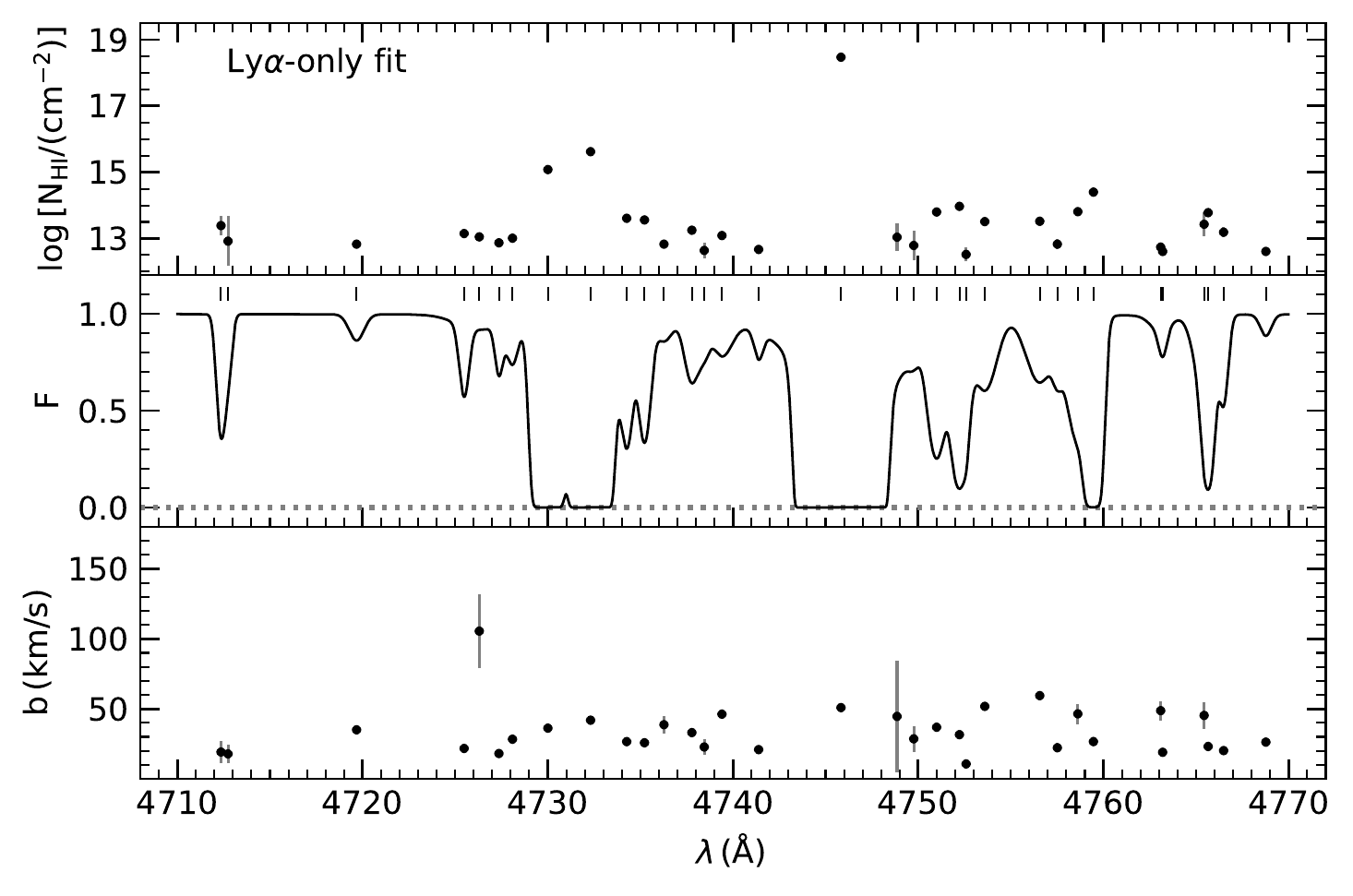}
\caption{
A portion of the reconstructed \lya\ forest spectrum, using $\NHI$ and $b$ parameters measured from QSO Q0636+6801 in \citet{Kim2021} either based on the Lyman series fit (left panels) and the \lya-only fit (right panels). The upper and lower panels show the $\NHI$ and $b$ measurements with their uncertainties, respectively, and the middle panels show the normalized flux with short vertical lines indicating the locations of the identified \lya\ absorbers.
}
\label{fig:mock_spec}
\end{figure*}

We refer interested readers to \citet{Kim2021} for details on the line analysis. As an illustration, Figure~\ref{fig:mock_spec} shows a portion of the reconstructed \lya\ forest spectrum from the line list of one quasar (Q0636+6801), with parameters based on the Lyman series fit (left) and the \lya-only fit (right). For each set, the reconstructed high-resolution spectrum is shown in the middle panel. As expected, there are no noticeable differences in the reconstructed spectra from the two sets of fitting parameters, since the fitting is done to reproduce absorption profiles. The small vertical lines in each middle panel mark the locations of individual absorbers, and the dots in the top and bottom panels are the values $\log\NHI$ and $b$ from Voigt fitting for these absorbers, as in the line list from \citet{Kim2021}. Note that the uncertainties in $\log\NHI$ and $b$ on the left panels are typically smaller, as not only \lya\ lines but also all available Lyman series lines are used in deriving these parameters. Lyman series lines of good signal-to-noise ratios also help resolve absorption structures, especially for saturated \lya\ absorptions. While this leads to small differences in the exact line lists in the left and right panels, it has little effect on the overall statistical properties of line parameters \citep{Kim2021}.

We will analyze the properties of absorbers from the line list \citep{Kim2021} and use them to constrain $T_0$ and $\gamma$ of the IGM thermal state.

%
%

\section{Constraining the Temperature-Density Relation from \lya\ Absorbers}
\label{sec:constraints}

We first analyze the overall distribution of the (logarithmic) neutral hydrogen column density $\logNHI$ and the Doppler parameter $b$ for the \lya\ absorbers in the \lya\ forests. Based on the distribution, we then employ an analytic model to constrain $T_0$ and $\gamma$, the two parameters describing the temperature-density relation of the IGM, using the $b$ cutoff as a function of $\logNHI$ determined from these \lya\ absorbers. Two methods are adopted for such constraints, as detailed below.  

\subsection{$b$--$\NHI$ Distribution and $b$ Cutoff}
\label{sec:overall}

\begin{figure*}
\flushleft 
\includegraphics[height=4.cm]{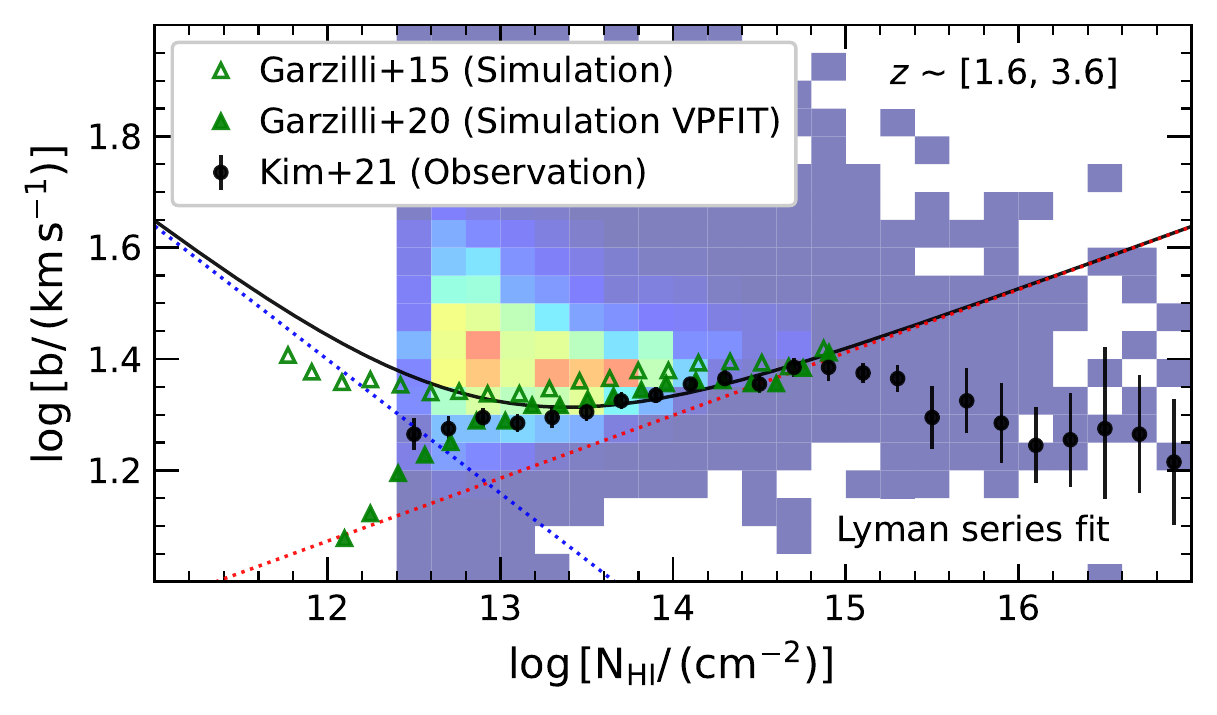}
\includegraphics[height=4.cm]{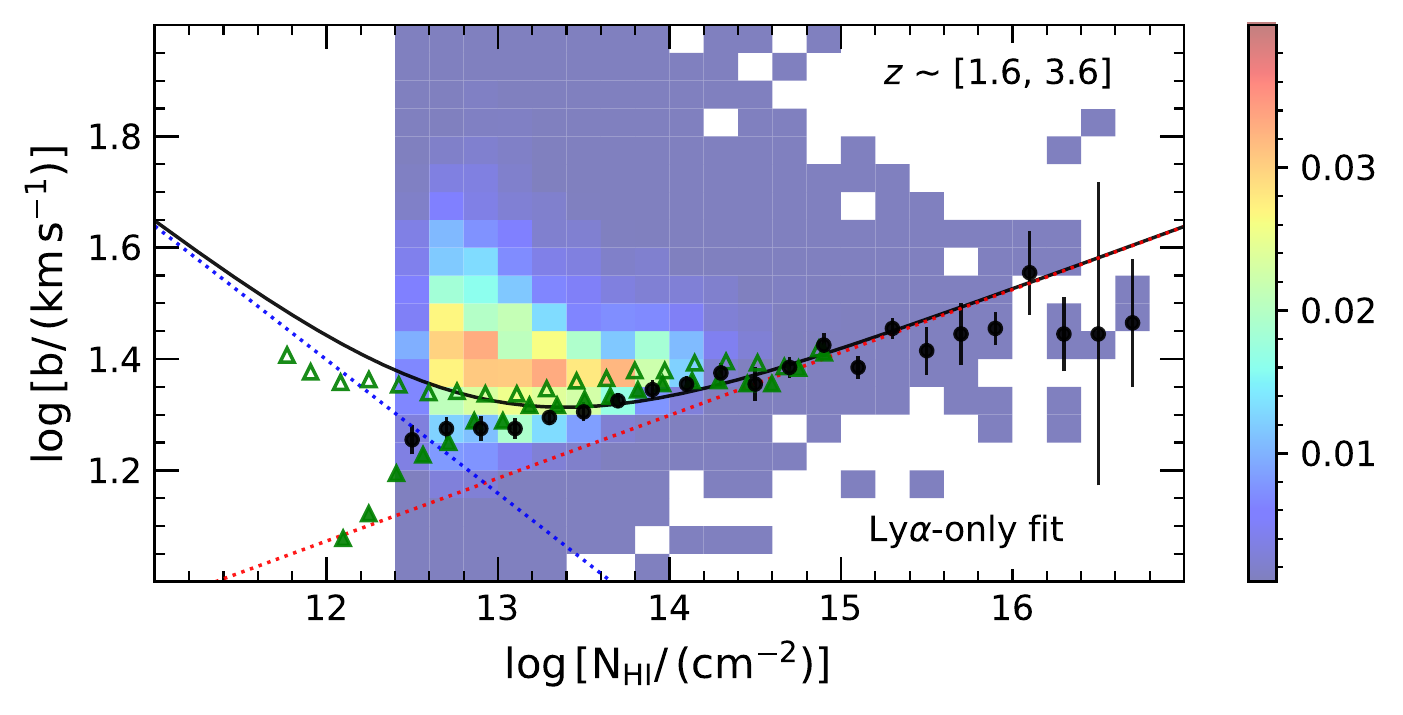}
\includegraphics[height=4.cm]{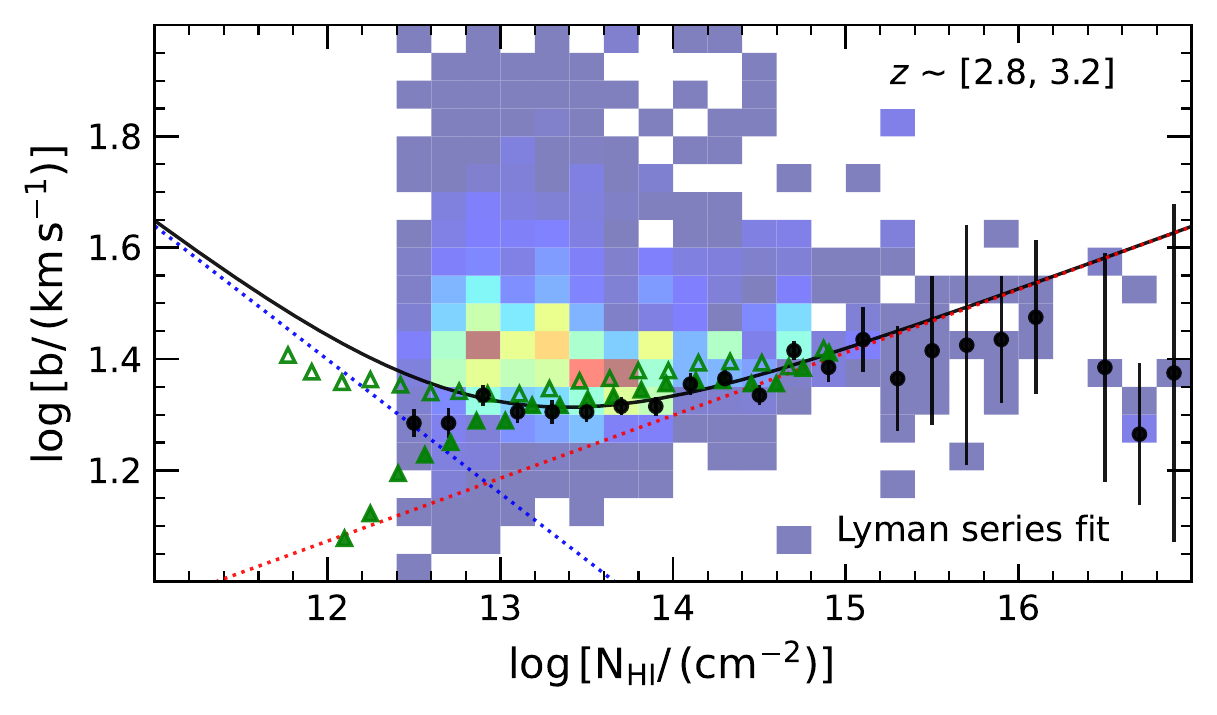}
\includegraphics[height=4.cm]{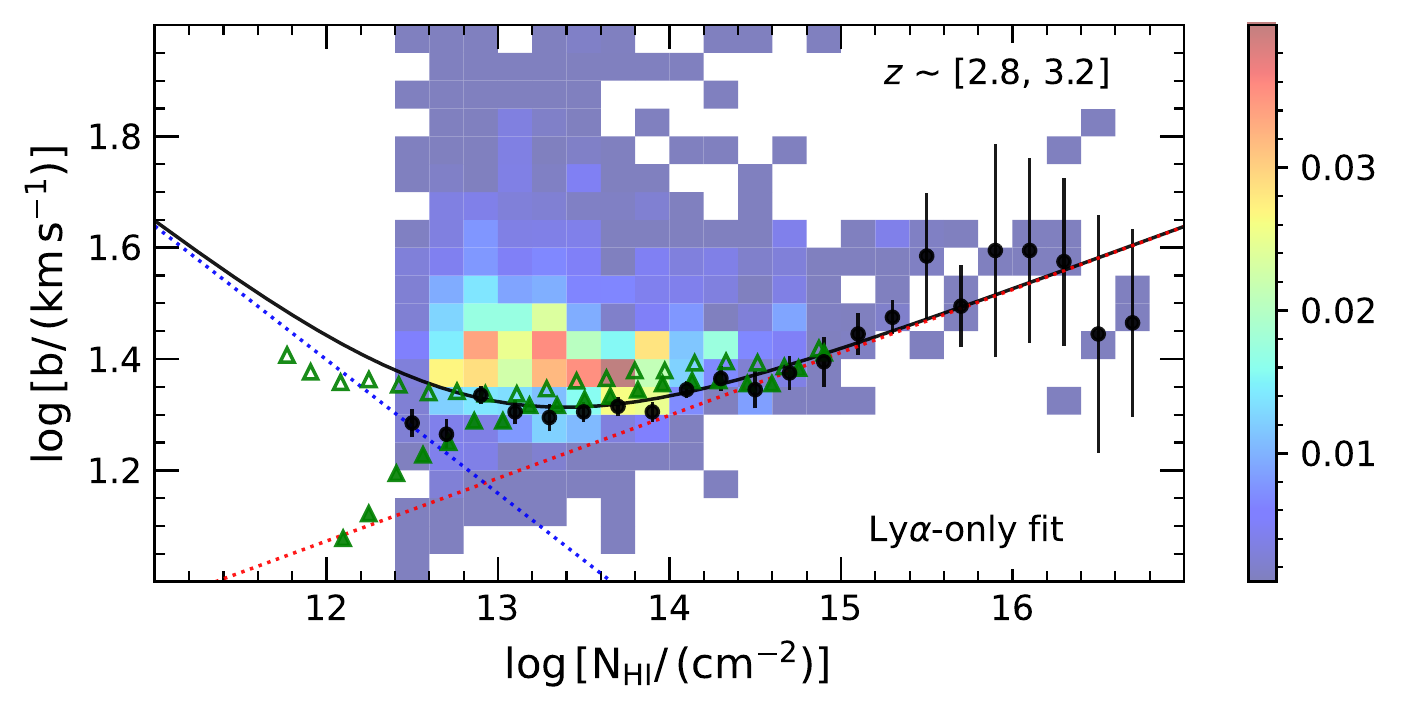}
\caption{
Distribution of $\NHI$ and $b$ parameters for \lya\ absorbers with redshift $z\in [1.6, 3.6]$ (upper panels) and $z\in [2.8, 3.2]$ (bottom panels), measured from the Lyman series fit (left panels) and the \lya-only fit (right panels). In each panel, the color-scale map shows the $b$--$\NHI$ distribution. The filled circles denote the lower 10th percentile of the $b$ distribution as a function of $\NHI$, with bootstrap error bars (see the text for more details). The open triangles show the lower 10th percentile relation of the $b$ parameters at $z\sim 3$ in \citet{Garzilli2015}, inferred from the hydrodynamic simulation, while the filled triangles correspond to the relation using $b$ parameters from Voigt fitting to the spectra in the hydrodynamic simulation in \citet{Garzilli2020}. The solid black curve is the $b$ cutoff relation from the analytic model developed in \citet{Garzilli2015}, decomposed into the contribution from the thermal broadening (red dotted line) and the Hubble broadening (blue dotted line (see the text). 
}
\label{fig:10th_logN_logb}
\end{figure*}

The color-scale maps in Figure~\ref{fig:10th_logN_logb} show the overall distribution of $\NHI$ and $b$ for the full absorber redshift range $z \in [1.6, 3.6]$ (top) and for $z \sim 3$ ($z \in [2.8, 3.2]$; bottom), based on the line lists from the Lyman series fit (left) and the \lya-only fit (right), respectively. 
We exclude lines with $b < 10\, \kms$ as they are most likely metal line contaminants or Voigt fit artifacts\footnote{
Theoretically, for thermal broadening at temperature $T$, the $b$ parameter for hydrogen follows $b=(2k_BT/m)^{1/2}=12.8(T/10^4{\rm K})^{1/2}~{\kms}$ and those for metals are much narrower. A threshold of $10~\kms$ to remove metal lines is a reasonable choice for typical IGM temperaures. Observationally, \citet{Hiss2018} visually inspect the absorption lines with $b < 10~\kms$ and identify them mainly as metal lines wrongly fit as \lya\ absorptions. In practice, the lines with $b< 10~\kms$ removed in our analysis are rare (at a percent level), which have virtually no effect on our results.
}. Lines with $b > 100\, \kms$ are also excluded as they have a larger contribution from turbulent broadening than from thermal broadening. These extremely broad lines are rare and discarding them does not affect any of our results.
To produce each map, we represent the likelihood of each pair of the $\logNHI$ and $\log b$ measurement as a bivariate Gaussian distribution using the measurement uncertainties and evaluate the sum of the likelihoods from all the absorbers in grid cells with $\Delta\logNHI=0.2$ and $\Delta\log b=0.01$. 
The results are shown with a coarser grid.

The $b$--$\NHI$ distributions in Figure~\ref{fig:10th_logN_logb} look similar to each other. The $b$ distribution peaks around 20--30 $\kms$, with a slightly higher value at the lower end of the column density. \lya\ absorption lines are broadened by both thermal motion and non-thermal broadening resulting from the combination of Hubble flow, peculiar velocities, and turbulence. In many applications \citep[e.g.,][]{Schaye1999,Schaye2000,Ricotti2000,McDonald2001,Rudie2012,Boera2014,Bolton2014,Garzilli2015,Hiss2018,Rorai2018,Telikova2021}, the narrowest \lya\ absorption lines in the \lya\ forests are identified and used to constrain the IGM thermal state, as the broadening of these lines is supposed to be purely thermal and the non-thermal broadening is negligible. 

The narrowest \lya\ absorption lines define the overall lower cutoff in the $b$ distribution as a function of $\NHI$. We perform such an analysis by computing the locus of the boundary of the lower 10th percentile in the $b$ distribution in each $\NHI$ bin. The black points in each panel of Figure~\ref{fig:10th_logN_logb} delineate such a cutoff boundary, with error bars estimated from bootstrap resampling the data points 100 times. The black solid curve is from an analytic model developed in \citet{Garzilli2015}, which describes the minimum line broadening (defined by the 10th percentile cutoff $b$) as the sum (in quadrature) of the thermal broadening (dotted red line) and the Hubble broadening (dotted blue line). 

The analytic curve is described by \citet{Garzilli2015} as
\begin{equation}
b^2=\frac{2k_B T_0}{m}N^{(\gamma-1)/\alpha} \left[1+0.75\left(\frac{f_J}{0.88}\right)N^{-1/\alpha}\right], \label{eqn:eq1}   
\end{equation}
with 
\begin{equation}
\alpha=1.76-0.26\gamma,
\end{equation}
\begin{equation}
N\equiv \left(\frac{\NHI}{N_0}\right)\left(\frac{T_0}{10^4{\rm K}}\right)^{0.26}\left(\frac{f_N}{0.3}\right)^{-1} \left(\frac{f_J}{0.88}\right)^{-1} , 
\end{equation}
and
\begin{eqnarray}
N_0=1.4\times 10^{13} {\rm cm^{-2}} \left(\frac{\Gamma}{10^{-12}{\rm s^{-1}}}\right)\left(\frac{1+z}{4}\right)^{9/2} \nonumber\\
\times \left(\frac{\Omega_b}{0.04825}\right)^2\left(\frac{\Omega_m}{0.307}\right)^{-1/2}\left(\frac{h}{0.6777}\right)^3,
\label{eqn:eq4}
\end{eqnarray}
where the temperature-density relation is parameterized as $T=T_0\Delta^{\gamma -1}$ \citep{Hui1997},
$\Gamma$ is the hydrogen photoionization rate, and $k_B$ and $m$ are the Boltzmann constant and the mass of hydrogen atom. The parameter $f_J$ describes the smoothing of the gas density profiles, which is the ratio of the filtering (smoothing) scale $\lambda_F$ to the Jeans length $\lambda_J$ \citep[e.g.,][]{Gnedin1998}. It is introduced when relating the neutral column density $\NHI$ to the neutral number density $n_{\rm HI}$ of hydrogen, $\NHI \propto n_{\rm HI} f_J\lambda_J$, with $f_N$ the proportionality factor in this relation \citep[][]{Schaye2001,Garzilli2015}.  At low and high column density, $b^2\propto \NHI^{(\gamma-2)/\alpha}$ and $b^2\propto \NHI^{(\gamma-1)/\alpha}$, and the value of $b$ is dominated by the Hubble broadening and the thermal broadening, respectively. 

The analytic model seems to match the cutoff boundary inferred from the line list at high column density. For $z\in [1.6, 3.6]$, this is above $\logNHI\sim 13.5$. While the agreement goes to $\logNHI\sim 16.5$ with the case of the \lya-only fit, the inferred boundary based on the Lyman series fit has a lower $b$ cutoff than the model above $\logNHI\sim 15$. A more appropriate comparison between our inference and the model is to limit the redshift range. As the model curve we plot is for $z \sim 3$, the bottom panels make a fair comparison. In this case, we find that our inferred boundary largely agrees with the model above $\logNHI\sim 13$, for line parameters from both the Lyman series fit and the \lya-only fit, with large uncertainties above $\logNHI\sim 15$. We note that the analytic model is in fact only accurate for $\logNHI\lesssim 15$ and overpedicts the lower $b$ cutoff for higher column density (see \citealt{Garzilli2015}), as at higher density additional effects from the balance between photoheating and radiative cooling need to be considered for a more accurate model. To be consistent, in constraining the IGM thermal states, we only use the data below $\logNHI= 15$, where the analytic model is valid.

Below $\logNHI\sim 13$, the data fall below the model curve. A possible cause is the incompleteness in the data -- for low column density absorption systems, those with high values of $b$ would show up as shallow absorption features in the quasar spectrum, which are hard to identify. In \citet{Garzilli2015}, a 10th percentile cutoff boundary in the $b$--$\NHI$ distribution for $z \sim 3$ absorbers is presented based on one OWLS simulation \citep{Schaye2010}, shown as the empty triangles in each panel of Figure~\ref{fig:10th_logN_logb}. The values of $\NHI$ and $b$ are directly computed from the simulation data. \citet{Garzilli2020} further provide the cutoff boundary from Voigt fitting to the simulated \lya\ forest spectra (see their Fig.A1), which resembles the procedure in analyzing observational data. This is shown as the solid triangles in Figure~\ref{fig:10th_logN_logb}. Compared to the case without Voigt fitting, the cutoff values of $b$ are lowered at the low column density end. That is, at fixed, low column density, Voigt fitting tends to miss shallow absorption lines with high values of $b$. It is encouraging that our inferred cutoff boundary (shown with black circles) is in good agreement with their simulation-based one from Voigt fitting, including the trend at column density below $\logNHI\sim 13$.

\begin{figure*}
    \centering
    \includegraphics[width=.85\textwidth]{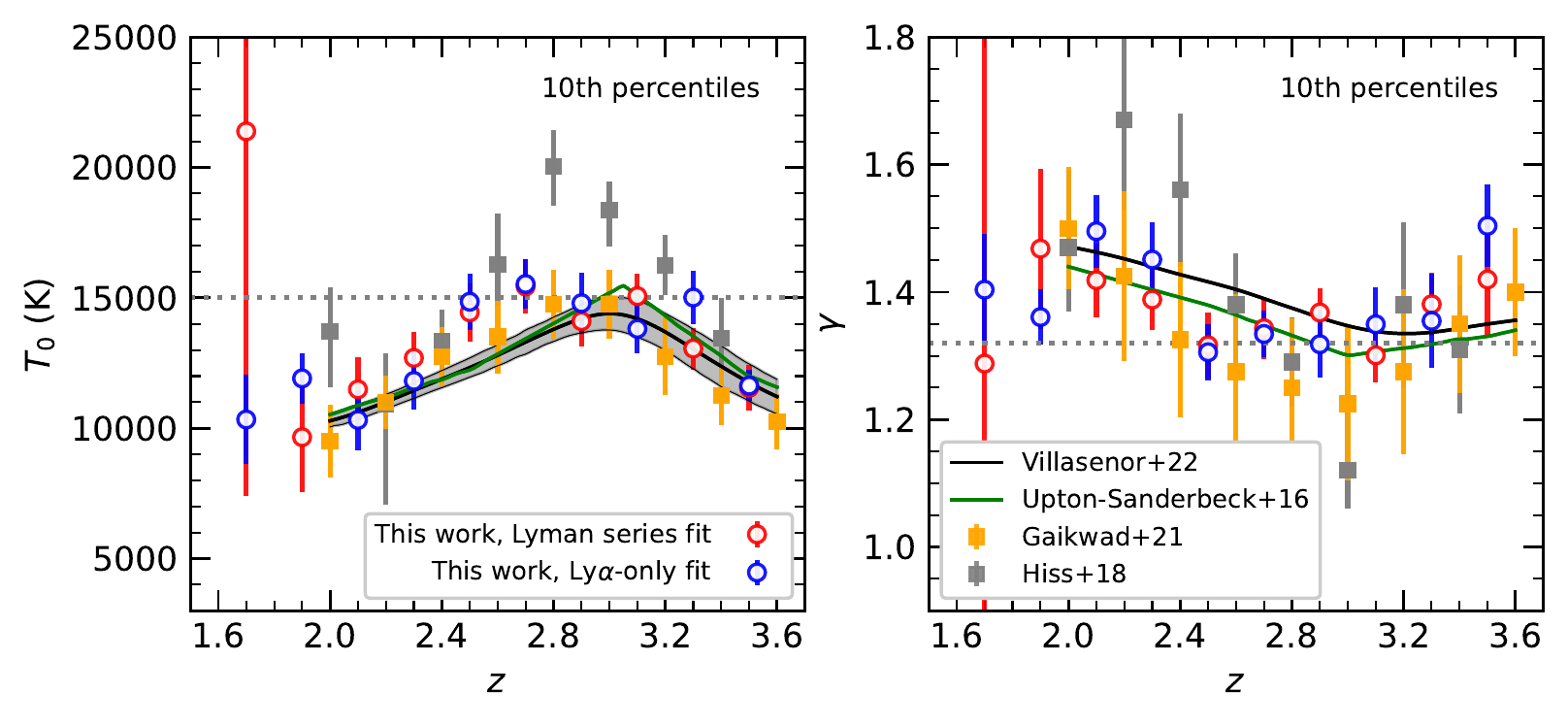}
    \includegraphics[width=.85\textwidth]{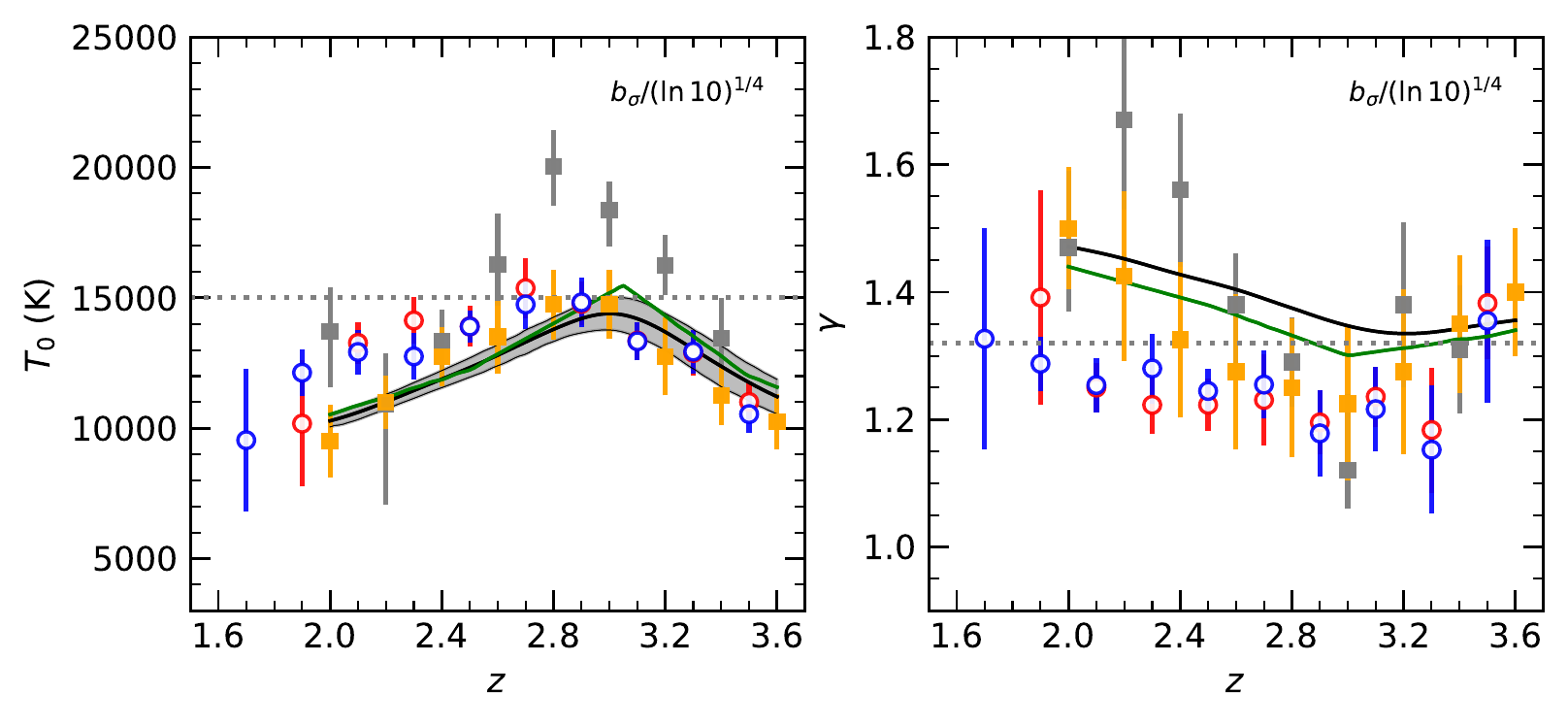}
    \caption{Constraints on $T_0$ and $\gamma$, the two parameters describing the temperature-density relation $T=T_0\Delta^{\gamma-1}$ in the IGM. Our results are shown as open circles, red (blue) circles correspond to \lya\ absorber parameters $\NHI$ and $b$ measured from the Lyman-series (\lya\-only) fit. In the top panels, our constraints are derived using the $b$ cutoff relation estimated from the data, while those in the bottom panels are from the $b$ cutoff relation inferred from fitting the $b$ distribution. See the text for details. In each panel, The gray squares are constraints from \citet{Hiss2018}, also from modeling the $b$ cutoff relation, but with a set of hydrodynamic simulations. The yellow squares are from \citet{Gaikwad2021}, constrained based on four different \lya\ forest flux statistics. The black curve (with the 95\% confidence interval shaded) corresponds to the constraints in \citet{Villasenor2022} by fiting the \lya\ forest power spectrum with a large set of hydrodynamic simulations. The green curve is the prediction in \citet{Upton2016} from a hydrodynamic simulation including the effect of \ion{He}{2} reionization. The horizontal line at $T_0=1.5\times 10^4$K or $\gamma=1.32$ is shown simply as a reference to aid the comparison.
    }
    \label{fig:T_gamma}
\end{figure*}

We make an attempt to use the analytic model developed in \citet{Garzilli2015}, i.e., equations~(\ref{eqn:eq1})--(\ref{eqn:eq4}), to constrain the redshift-dependent $T_0$ and $\gamma$, the two parameters describing the temperature-density relation, $T=T_0\Delta^{\gamma-1}$. We first fix $f_N$, $f_J$, and $\Gamma$ at their fiducial values as in equations~(\ref{eqn:eq1})--(\ref{eqn:eq4}) and will discuss possible systematic effects introduced by adopting these values. The cosmological parameters ($\Omega_b$, $\Omega_m$, and $h$) are also fixed at their fiducial values, which are consistent with the Planck constraints \citep{Planck2020}.

The observed \lya\ absorbers over the range of $1.6<z<3.6$ are divided into 10 redshift bins. The numbers of absorbers in the ten redshift bins from the Lyman series fit (\lya-only fit) are 35 (272), 260 (879), 1183 (1178), 941 (930), 830 (948), 604 (715), 669 (650), 524 (523), 296 (288), 273 (255), respectively. We use lines with $\log\NHI$ in the range of [13, 15], since lines with $\log\NHI < 13$ are incomplete for weak lines and those with $\log\NHI > 15$ are too saturated to have reliable Voigt parameter measurements with the \lya-only fit. 

We use the 10th percentile cutoff in $b$ for the parameter constraints, as the analytic model is tuned for such a cutoff threshold. Two methods are applied to estimate the 10th percentile cutoff in $b$ in each $\NHI$ bin, a non-parametric method that directly measures the 10th percentile from the cumulative $b$ distribution and a parametric method from fitting the full $b$ distribution, as detailed in the following two subsections.

\subsection{$T_0$--$\gamma$ Constraints from the 10th percentile $b$ cutoff estimated from a non-parametric method}
\label{sec:T0gamma_10th}

We first present the results based on the 10th percentile cutoff profile estimated using a non-parametric method. Similar to those done in Section~\ref{sec:overall} and in Figure~\ref{fig:10th_logN_logb}, at a given redshift, in each $\log\NHI$ bin, we derive the 10th percentile cutoff boundary of $b$ by computing the cumulative distribution function of $b$, where each $b$ measurement is taken as a Gaussian distribution with the standard deviation set by the observational uncertainty. The uncertainty in the 10th percentile locus is estimated through bootstrap resampling the data points 100 times. The analytic model is then applied to constrain $T_0$ and $\gamma$. 

The results are shown in the top two panels of Figure~\ref{fig:T_gamma} (labeled as ``10th percentiles''), and $T_0$ and $\gamma$ constraints as a function of $z$ are found in Table~\ref{table_T0gamma}. The red and blue points are based on $b$-$\NHI$ parameters measured through the Lyman series fit and the \lya-only fit, respectively. They appear to be consistent with each other, typically within 1$\sigma$. Compared to those based on the \lya-only fit, those based on the Lyman series fit have larger uncertainties at lower redshifts, since the number of available sightlines for the Lyman series fit is smaller.

Both the values of $T_0$ and $\gamma$ show a clear trend with redshift, with a transition around $z\sim 2.8$. The temperature $T_0$ at mean density increases from $\sim 10^4$K at $z\sim 1.7$ to $\sim 1.55\times 10^4$K at $z\sim 2.7$, then decreases towards higher redshifts, reaching $\sim 1.15\times 10^4$K at $z\sim 3.5$. The $T_0$ value at $z\sim 3.1$ for the Lyman series fit case and that at $z\sim 3.3$ for the \lya-only fit case deviate the trend of decreasing with increasing redshifts, with $T_0\sim$1.4--1.5$\times 10^4$K, but they are consistent with being fluctuations. For the values of $\gamma$, the broad trend appears to be decreasing from $\gamma \sim$1.4--1.5 to 1.3 in the redshift range of 1.6 to 2.8--2.9 and then flattened (or maybe slightly increasing) towards higher redshifts. The transitions seen in $T_0$ and $\gamma$ around $z\sim$2.7--2.9 are signatures of \ion{He}{2} reionization \citep[e.g.,][]{Upton2016,Worseck2019,Villasenor2022}. 
Photons ionizing \ion{He}{2} heat the IGM, and the temperature $T_0$ climbs up, reaches a peak, and then drops when adiabatic expansion starts to dominate the temperature evolution. Heating the IGM makes $\gamma$ decrease (e.g., $\gamma$ would become unity if the IGM is heated to be isothermal) and then it increases when the effect of adiabatic expansion kicks in. 

The temperature-density relation has been observationally constrained with various methods. 
Early constraints show large scatters and have large uncertainties. We choose to compare to a few recent constraints.
As a comparison, the gray squares in the top panels of Figure~\ref{fig:T_gamma} are from \citet{Hiss2018}. They are also constrained through the low-$b$ boundary of the $b$-$\NHI$ distribution with high-resolution spectra, but by comparing to a set of hydrodynamic simulations. The overall trend is similar to ours, e.g., a peak in $T_0$ at $z\sim 2.8$. The variation amplitude in $T_0$ from \citet{Hiss2018} appears higher than ours -- while the $T_0$ values at the low- and high-redshift ends are consistent with ours, their peak $T_0$ value is much higher, $\sim 2\times 10^4$K, $\sim$25\% higher than ours ($1.55\times 10^4$K). Similarly, the value of $\gamma$ from their constraints has a steeper drop from $z\sim 2$ to $z\sim 2.9$ (with larger uncertainties though). 

\citet{Villasenor2022} constrain the temperature-density relation and the evolution of the ionization rate by fitting the \lya\ forest power spectrum from high-resolution spectroscopic observations using a large set of hydrodynamic simulations. The black curves in the top panels show $T_0$ and $\gamma$ constraints from their bestfit model, with the shaded bands representing the 1$\sigma$ uncertainty (very narrow in the $\gamma$ constraints). Our results agree well with theirs in terms of the variation amplitude of $T_0$, while the peak in our results occurs at a lower redshift ($z\sim 2.7$) than theirs ($z\sim 3.0$). The trends in $\gamma$ are also similar, with our results showing a slightly lower (higher) $\gamma$ at low (high) redshifts. Given the uncertainties in our inferred $\gamma$ values, our results are consistent with theirs. 

The yellow data points are constraints from \citet{Gaikwad2021}, based on four different flux distribution statistics of \lya\ forests in high-resolution and high-S/N quasar spectra. Accounting for the uncertainties, our results show a good agreement with theirs.

The green curve in each panel represents the prediction from a hydrodynamic simulation in \citet{Upton2016}, which includes the effect of \ion{He}{2} reionization. It almost falls on top of the constraints in \citet{Villasenor2022} for $T_0$. It appears to be slightly lower in $\gamma$, but with a quite similar trend. The broad features predicted from this hydrodynamic simulation are similar to our results, except for the small shift in the redshift of the peak $T_0$.

\subsection{$T_0$--$\gamma$ Constraints from the 10th percentile $b$ cutoff estimated from a parametric fit to the $b$ distribution}
\label{sec:T0gamma_dNdb}

Estimating the $b$ cutoff directly from the cumulative $b$ distribution, while straightforward, can have limitations. First, the IGM thermal state impacts all the lines, not just the narrowest lines. Therefore, by restricting  the use of the data in the tail of the distribution near the cutoff, this approach throws away information, which can significantly reduce the sensitivity to the IGM thermal state. Second, in practice, determining the location of the cutoff is vulnerable to systematic effects, such as contamination from unidentified metal lines or misidentified metal lines as \HI\ and noise. 

To overcome these limitations, we develop an approach to infer the 10th percentile $b$ cutoff by a parametric fit to the full $b$ distribution in each $\NHI$ bin. To describe the $b$ distribution, we adopt the functional form suggested by \cite{Hui1999}, derived based on the Gaussian random density and velocity field. It is a single-parameter distribution function,
\begin{equation}
    \frac{dN}{db} \propto \frac{b_{\sigma}^4}{b^5}\exp\left(-\frac{b_{\sigma}^4}{b^4}\right).\label{eqn:Hui_Rutledge_function}
\end{equation}
Such a distribution function naturally explains the salient features of the observed $b$ distribution: a sharp low-$b$ cutoff, corresponding to narrow and high amplitude absorptions (statistically rare, related to the tail of the Gaussian distribution), and a long power-law tail toward high $b$, coming from broad and low-amplitude absorptions. The parameter $b_\sigma$ marks the transition from the exponential cutoff to the power-law part. With this distribution, the $b$ value for the lower 10 percentile cutoff is $b_\sigma/(\ln 10)^{1/4}$.

For every redshift bin, in each $\NHI$ bin, with the observed values of $b$, we derive the constraints on $b_\sigma$ using the maximum likelihood method\footnote{We adopt the publicly available Python Package \texttt{kafe2} (\url{https://github.com/PhiLFitters/kafe2}) to perform the maximum likelihood estimation.}.
The inferred values of the 10th percentile cutoff $b_\sigma/(\ln 10)^{1/4}$ as a function of $\NHI$ are used to constrain $T_0$ and $\gamma$, as in Section~\ref{sec:T0gamma_10th}. The results are shown in the bottom panels of Figure~\ref{fig:T_gamma} (labeled as ``$b_\sigma/(\ln 10)^{1/4}$'').

The constraints on $T_0$ are similar to those inferred from using the 10th percentile $b$ cutoff estimated from the non-parametric method. The peak is around $z\sim 2.8$ with a value of $\sim 1.5\times 10^4$K. The fluctuation in the trend with redshift is reduced, as expected, given that the 10th percentile boundary is from fitting the overall $b$ distribution. For $\gamma$, the trend with redshift is also similar to that based on the non-parametric $b$ cutoff estimate, but the amplitude appears to be lower. The systematic trend may reflect the fact that the analytic model is tuned for the first method, not the second one. However, for most $\gamma$ values the systematic shifts are within 2$\sigma$ with the data we use in our analysis. 

As a whole, our derived constraints on $T_0$ and $\gamma$ based on two methods of estimating the $b$ cutoff profile broadly agree with each other. They also appear to be consistent with the results inferred by \citet{Gaikwad2021}. While the main difference in our two types of constraints lies in the amplitude of $\gamma$, both of them appear to be around the \citet{Gaikwad2021} values, typically within 1$\sigma$.

We present the results here as the marginalized constraints on $T_0$ and $\gamma$, respectively. For completeness, the full constraints with both methods of deriving the $b$ cutoff are shown in the Appendix~\ref{sec:appendix}, where we also provide Table~\ref{table_T0gamma} for the $T_0$ and $\gamma$ constraints.

\subsection{Sensitivity of the $T_0$--$\gamma$ constraints on parameters in the analytic model}
\label{sec:sensitivity}

\begin{figure*}
    \centering
    \includegraphics[width=.85\textwidth]{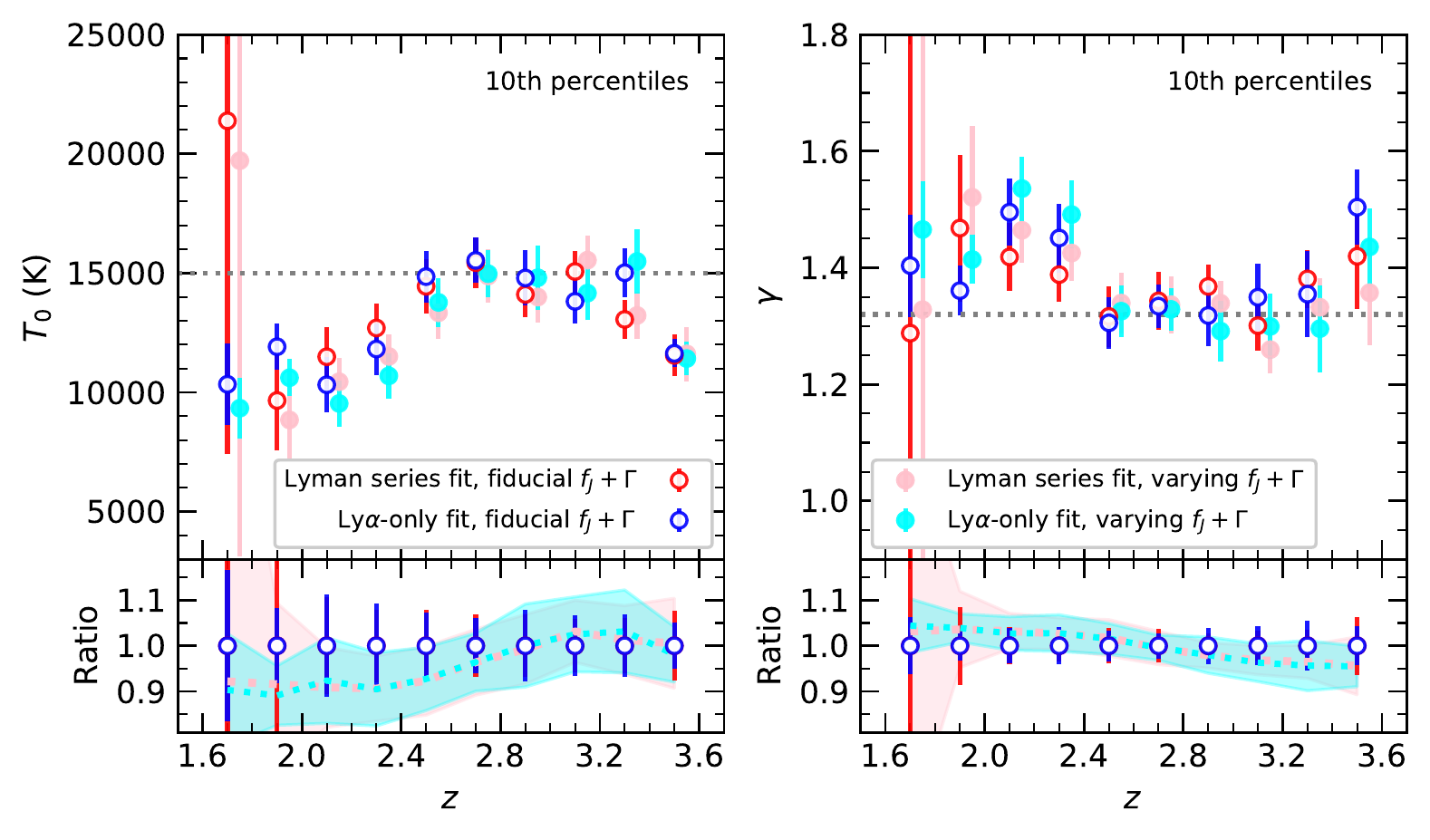}
    \includegraphics[width=.85\textwidth]{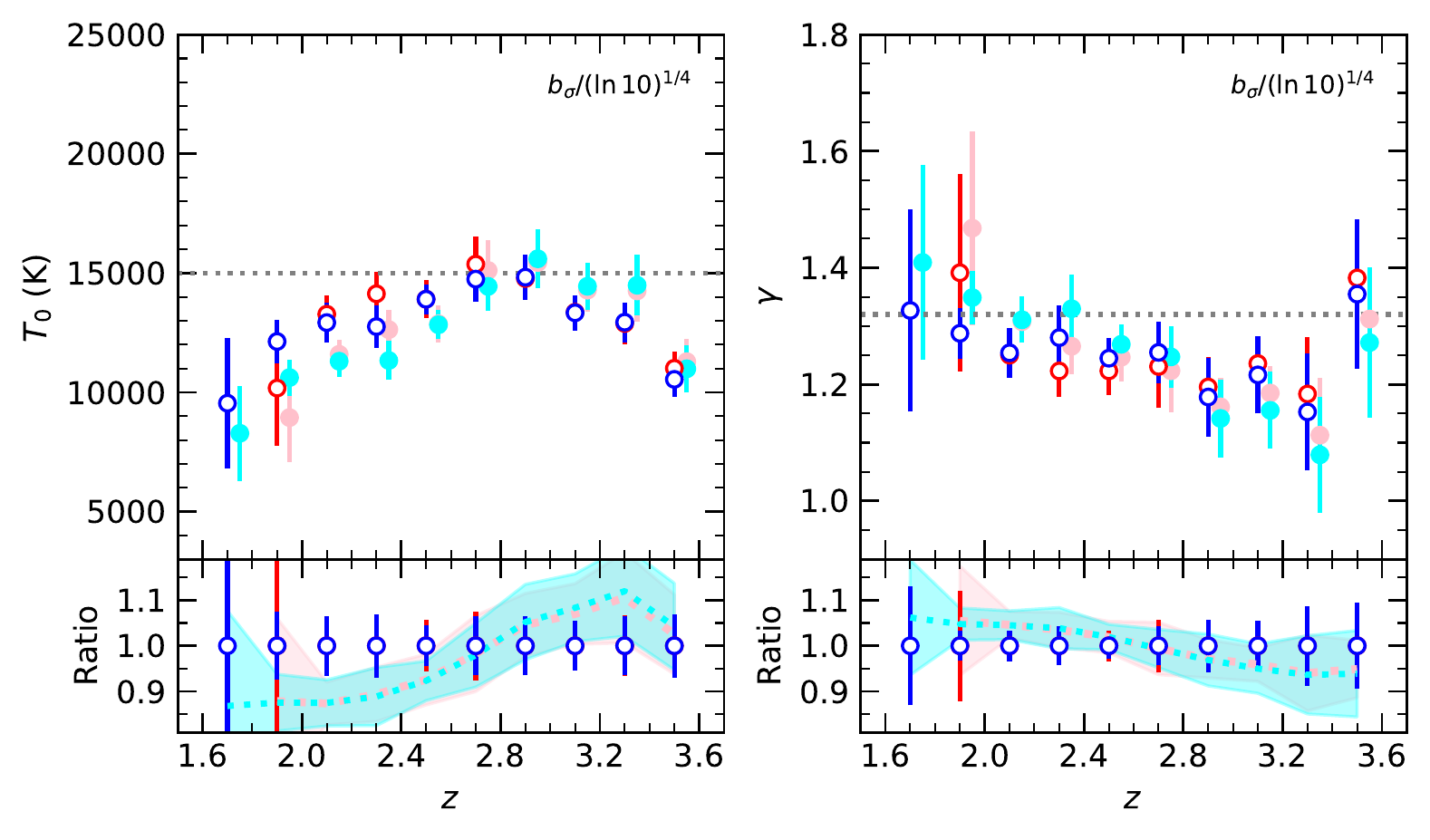}
    \caption{Similar to Fig.~\ref{fig:T_gamma}, but our constraints are compared to those with evolving model parameters. The red and blue open circles are the same as in Fig.~\ref{fig:T_gamma}, where $f_J$ and $\Gamma$ parameters in the analytic model are fixed at their fiducial values. The pink and cyan filled circles are constraints when allowing $f_J$ and $\Gamma$ to evolve with redshift. See the text for details. For clarity, the pink and cyan points are shifted by $\Delta z=0.05$. The ratios of the latter constraints to the former ones are shown in the small panels, with the shaded regions representing the 1$\sigma$ uncertainties. 
    }
    \label{fig:Ratio}
\end{figure*}

In obtaining the constraints on $T_0$ and $\gamma$ in Sections~\ref{sec:T0gamma_10th} and \ref{sec:T0gamma_dNdb}, we have fixed the model parameters $f_N$, $f_J$, and $\Gamma$ to their fiducial values in equations~(\ref{eqn:eq1})--(\ref{eqn:eq4}). That is, we neglect their dependence on redshift. One may worry that this could introduce systematic uncertainties in the inferred $T_0$ and $\gamma$.

We can study the sensitivity of the $T_0$ and $\gamma$ constraints on these model parameters by considering the low-$\NHI$ and high-$\NHI$ limit in the model. At low $\NHI$, where the Hubble broadening dominates, the cutoff in $b$ is approximately
\begin{equation}
    b^2\propto T_0^{1+\frac{0.26(\gamma-2)}{\alpha}} f_J^{1-\frac{\gamma-2}{\alpha}} (f_N\Gamma)^{-\frac{\gamma-2}{\alpha}} \NHI^{\frac{\gamma-2}{\alpha}},
    \label{eqn:loNHI}
\end{equation}
and at high $\NHI$, where the thermal broadening dominates, the approximation becomes
\begin{equation}
    b^2\propto T_0^{1+\frac{0.26(\gamma-1)}{\alpha}} (f_J f_N\Gamma)^{-\frac{\gamma-1}{\alpha}} \NHI^{\frac{\gamma-1}{\alpha}},
    \label{eqn:hiNHI}
\end{equation}
where $\alpha=1.76-0.26\gamma$. In each expression, the power on $T_0$ is insensitive to $\gamma$, while the power on $\NHI$ is sensitive to $\gamma$. That is, with the $\NHI$-dependent $b$ cutoff profile, it is mainly the amplitude that determines $T_0$ and the shape that constrains $\gamma$. 

Our constraints are from using the data at $\log\NHI>13$, mostly in the high-$\NHI$ regime. If we take $\gamma\sim 1.35$, equation~(\ref{eqn:hiNHI}) becomes $b^2\propto T_0^{1.06}(f_J f_N \Gamma)^{-0.25}\NHI^{0.25}$. A 20\% change in any combination of the $f_J$, $f_N$, and $\Gamma$ parameters only leads to $\sim$5\% change in the $T_0$ constraints. Such a change is well within the uncertainty in the constraints shown in Figure~\ref{fig:T_gamma}. Since the above change in $f_N$, $f_J$, and $\Gamma$ can be largely absorbed into the $T_0$ constraints, the constraints on $\gamma$ from the dependence on $\NHI$ would not be affected much. 

Around $\log \NHI\sim 13$, both the Hubble broadening and thermal broadening contribute to the $b$ cutoff (see the curves in Fig.~\ref{fig:10th_logN_logb}), and we perform further tests by varying each of $f_J$, $f_N$, and $\Gamma$ by 20\% in the model to fit the data. We find that varying $f_N$ and/or $\Gamma$ only leads to a couple of percent effects on the $T_0$ and $\gamma$ constraints. The effect of varying $f_J$ is larger, expected from equation~(\ref{eqn:loNHI}), which at low $\NHI$ becomes $b^2\propto T_0^{0.88}f_J^{1.46} (f_N \Gamma)^{0.46}\NHI^{-0.46}$ for $\gamma\sim 1.35$. The corresponding variation in the $T_0$ constraints ranges from $\sim 6$\% (low redshifts) to $\sim$10--15\% (high redshifts). In high redshift bins, these are about 2--3$\sigma$ changes in the value of $T_0$. For $\gamma$, the change in the constraints ranges from $\sim$2.5\% to $\sim$6\%, still well within the uncertainty. 

The above tests show that our results are not significantly affected by our choice of fixing $f_J$, $f_N$, and $\Gamma$ to their fiducial values. As an additional test, we also infer the constraints by removing data around $\log \NHI\sim 13$ and limiting the data to $13.5<\logNHI<15$, and the results remain consistent with our original ones.

While the above tests to the sensitivity are general, they do not reflect the expected redshift dependence of the model parameters. For a more realistic assessment of the potential systematics, we turn to simple models of these parameters. The parameter $f_N$, which is the proportionality coefficient relating $\NHI$ and the product of the number density $n_{\rm HI}$ and the filtering scale $\lambda_F=f_J\lambda_J$, is expected to be insensitive to redshift. The filtering (smoothing) scale $\lambda_F$ at a given epoch depends on the history of the Jeans length $\lambda_J$, i.e., on the thermal history of the IGM. Therefore, we expect $f_J$ to evolve with redshift. The photoionization rate $\Gamma$ is expected to depend on redshift, as the ionizing photons come from the evolving populations of star-forming galaxies and quasars. We perform further tests on the $T_0$-$\gamma$ constraints by modeling $f_J(z)$ and $\Gamma(z)$.

\citet{Gnedin1998} derive an analytic expression of $\lambda_F$ in linear theory,
\begin{equation}
\begin{split}
\lambda_F^2(t)= \frac{1}{D_+(t)}\int_0^t dt^\prime \lambda_J^2(t^\prime) a^2(t^\prime)  \times \\
\left[ \ddot{D}_+(t^\prime)+2H(t^\prime)\dot{D}_+(t^\prime)\right] \int_{t^\prime}^t \frac{dt''}{a^2(t'')},
\end{split}
\label{eqn:jeans}
\end{equation}
where $a(t)$ is the scale factor, $H(t)=\dot{a}/a$ is the Hubble parameter, $D_+(t)$ is the linear growth factor, and $\lambda^2_J(t)\propto a c_s^2/(G\rho_{m,0})$ is the square of the (comoving) Jeans length, with $c_s \propto T_0^{1/2}$ the sound speed and $\rho_{m,0}$ the comoving matter density. At high redshifts appropriate for our analysis here, the Einstein-de Sitter cosmology is a good approximation, with $D_+(t) \propto a(t)\propto t^{2/3}$. The expression then simplifies to
\begin{equation}
    \lambda_F^2(a)=\frac{3}{a}\int_0^a da'  \lambda_J^2(a')\left[1-\left(\frac{a'}{a}\right)^{1/2}\right].
\label{eqn:JeansSimple}
\end{equation}
To test the potential effect of the evolution of $f_J(z)=\lambda_F(z)/\lambda_J(z)$ on our constraints, we compute $f_J(z)$ by adopting an IGM temperature evolution similar in shape to that in \citet{Villasenor2022}, starting from $T_0\sim 0$K at $z\gtrsim 7$ and increasing to $T_0\gtrsim 10^4K$ at lower redshifts with two bumps at $z\sim 6$ and $z\sim 3$ caused by \ion{H}{1} and \ion{He}{2} reionization, respectively. The values of $f_J$ are scaled such that $f_J(z=3)=0.88$ to match the fiducial value tuned in \citet{Garzilli2015}. The resultant $f_J(z)$ is $\sim$0.9 for $2.9<z<3.6$ and ramps up towards lower redshifts to a value of $\sim$1.27 at $z\sim 1.6$.

In the redshift range of interest here, the empirically measured hydrogen photoionization rate $\Gamma$ only show a mild evolution \citep[e.g.,][]{Becker2007,Becker2013,Villasenor2022}. We model the evolution to be $\Gamma(z)=10^{-12}{\rm s^{-1}} [(1+z)/2.6]^{-1}$, consistent with the model in \citet{HM12} and those empirical measurements with a steeper evolution.

The test results with the evolving $f_J$ and $\Gamma$ are shown in Figure~\ref{fig:Ratio}, in comparison with the fiducial results. For the constraints using the 10th percentile $b$ cutoff estimated non-parametrically (top panels), $T_0$ appears to be slightly lower at lower redshifts, before reaching the peak. The value of $\gamma$ is slightly higher at lower redshifts and lower at higher redshifts. Each lower small panel shows the ratio of the constraints to those from the fiducial ones, and all the changes in the constraints caused by the evolving $f_J$ and $\Gamma$ are well within the 1$\sigma$ uncertainty.

For the results using the $b$ cutoff determined by a parametric fit to the $b$ distribution (bottom panels), the trends are similar to those in the top panels, but with changes of larger amplitude. For $\gamma$, the changes are still well within the 1$\sigma$ uncertainty. For $T_0$, most of the changes are also within 1$\sigma$ and others are within 1.5$\sigma$ (if accounting for uncertainties in the constraints with both fixed and varying $f_J$ and $\Gamma$). For both methods, adopting the evolving $f_J$ and $\Gamma$ leads to constraints more in line with those from \citet{Gaikwad2021} and \citet{Villasenor2022} at lower redshifts. 

As a whole, the tests demonstrate that fixing the parameters in the analytic model to their fiducial values does not introduce significant systematic trends in the $T_0$-$\gamma$ constraints with the data we use. The amplitude and shape of the column-density-dependent $b$ cutoff profile of \lya\ absorbers at $\log \NHI \in$ [13, 15] enable robust constraints on the temperature-density relation of the IGM around $z\sim 3$ within the framework of the analytic model.

%
%
\section{Summary and Discussion}
\label{sec:summary}

Based on the distribution of the neutral hydrogen column density $\NHI$ and Doppler $b$ parameter measurements of $1.6<z<3.6$ \lya\ absorbers in the \lya\ forest regions of high-resolution and high-S/N quasar spectra, we employ an analytic model to constrain $T_0$ and $\gamma$, the two parameters describing the temperature-density relation of the IGM, $T=T_0\Delta^{\gamma-1}$. The constraints come from the $\NHI$-dependent low $b$ cutoff, contributed by \lya\ absorbers dominated by thermal broadening. The IGM temperature $T_0$ at the mean density shows a peak of $\sim 1.5\times 10^4$K at $z\sim $2.7--2.9 and drops to $\sim 10^4$K at the lower and higher end of the redshift range. The index $\gamma$ reaches a minimum around $z\sim 3$. The evolution in both parameters signals that \ion{He}{2} reionization finishes around $z\sim 3$.

The low $b$ cutoff profile as a function of $\NHI$ is obtained using two methods. The first one is a non-parametric method. With the measured values of $\NHI$ and $b$ and their uncertainties, in each $\NHI$ bin, we compute the cumulative distribution of the measured $b$ parameter to find the cutoff value corresponding to the lower 10th percentile and the uncertainty in the cutoff value is obtained through bootstrapping. The second method is a parametric one. In each $\NHI$ bin, we fit the $b$ distribution with an analytic function \citep[][]{Hui1999} using a maximum likelihood method to infer the 10th-percentile $b$ cutoff value. The analytic model developed in \citet{Garzilli2015} is then applied to model these $b$ cutoff profiles to obtain the $T_0$-$\gamma$ constraints. For $T_0$, using the $b$ cutoff profiles estimated from the two methods leads to similar constraints. For $\gamma$, using the $b$ cutoff profile from fitting the $b$ distribution results in lower values of $\gamma$ than that using the non-parametrically inferred $b$ cutoff. This may result from the fact that the analytic model is effectively calibrated with the first method. Nevertheless, the constraints of $\gamma$ with $b$ cutoff profiles from the two methods are consistent within 2$\sigma$ in most redshift bins. In obtaining the constraints, we use $\NHI$ and $b$ parameters measured from fitting the Lyman series lines and from only fitting the \lya\ lines, respectively, and the results agree with each other.

Our results are in line with some recent $T_0$--$\gamma$ constraints from completely different approaches. Those include \citet{Gaikwad2021}, who measure the IGM thermal state by using four different flux statistics in the \lya\ forest regions of high-resolution and high-S/N quasar spectra and by using a code developed to efficiently construct models with a wide range of IGM thermal and ionization histories without running full hydrodynamic simulations. Our results also agree with those in \citet{Villasenor2022}, where the constraints on the IGM thermal and ionization history are obtained from modeling the one-dimensional \lya\ forest power spectrum with a massive suite of more than 400 high-resolution cosmological hydrodynamic simulations. These nontrivial agreements with other work of different approaches suggest that the analytic model we adopt not only correctly captures the main physics in the low $b$ cutoff but also is reasonably calibrated.

There are a few model parameters in the analytic model: $f_J$ relates the Jeans length to the filtering (smoothing) scale, $f_N$ is the proportional coefficient in determining the neutral hydrogen column density from the neutral hydrogen number density and the filtering scale, and $\Gamma$ is the hydrogen photoionization rate. At high $\NHI$ that we mainly use for the $T_0$--$\gamma$ constraints, the $b$ cutoff profiles and hence the constraints are insensitive to these parameters, e.g., with $b^2$ approximately depending on $(f_J f_N \Gamma)^{-0.25}$. We further test the sensitivity by adopting an evolving $f_J$ factor from linear theory and an assumed thermal evolution of the IGM and an observationally and theoretically motivated $\Gamma$ evolution, and we find no significant changes in the constraints. That is, adopting the fiducial values of the model parameters result in no significant systematic trend in the $T_0$--$\gamma$ constraints within the framework of the analytic model and with the observational data in our analysis. 

The analytic model, with its current functional form, however, could still have systematic uncertainties in that it may not perfectly fit the results from the hydrodynamic simulations. \citet{Garzilli2020} have discussed the possible improvements to the model. While directly using simulations to perform parameter constraints \citep[e.g.,][]{Villasenor2022} is a route to largely reduce systematic uncertainties, it would still be useful to calibrate an analytic model with a set of hydrodynamic simulations at different output redshifts. To be self-consistent,  model parameters like $f_J$ and $\Gamma$ should encode the dependence on the IGM's thermal and ionization history. The model can also be calibrated to accommodate different ways of inferring the $b$ cutoff profile, as well as different percentile thresholds for defining the cutoff \citep[e.g.,][]{Garzilli2020}. Such a model would have the advantage of being computationally efficient and can be easily applied to model observed \lya\ absorbers to learn about the physical properties of the IGM. 

At $z\gtrsim 3$, when \ion{He}{2} reionization is not complete, a large number of sightlines are needed to fully probe the IGM state with patchy \ion{He}{2} reionization. The sample of the 24 high-resolution and high-S/N quasar spectra used in our analysis may still have appreciable cosmic variance (more exactly sample variance) effects, and the uncertainties in our $T_0$ and $\gamma$ may have been underestimated. In fact, this is true for the constraints in most work. A large sample of \lya\ forest absorbers from high-resolution and high-S/N quasar spectra is desired to probe the IGM state and \ion{He}{2} reionization, which would help tighten the constraints on $T_0$ and $\gamma$ and also make it possible to constrain quantities like $f_J$ and $\Gamma$.

\begin{acknowledgements}

This work is supported by National Key R\&D Program of China (grant No. 2018YFA0404503). L.Y. gratefully acknowledges the support of China Scholarship Council (No. 201804910563) and the hospitality of the Department of Physics and Astronomy at the University of Utah during her visit. Z.Z. is supported by NSF grant AST-2007499. The support and resources from the Center for High Performance Computing at the University of Utah are gratefully acknowledged.

\end{acknowledgements}

%
%

\appendix
\section{}
\label{sec:appendix}

\begin{figure*}
    \centering
    \includegraphics[height=7cm]{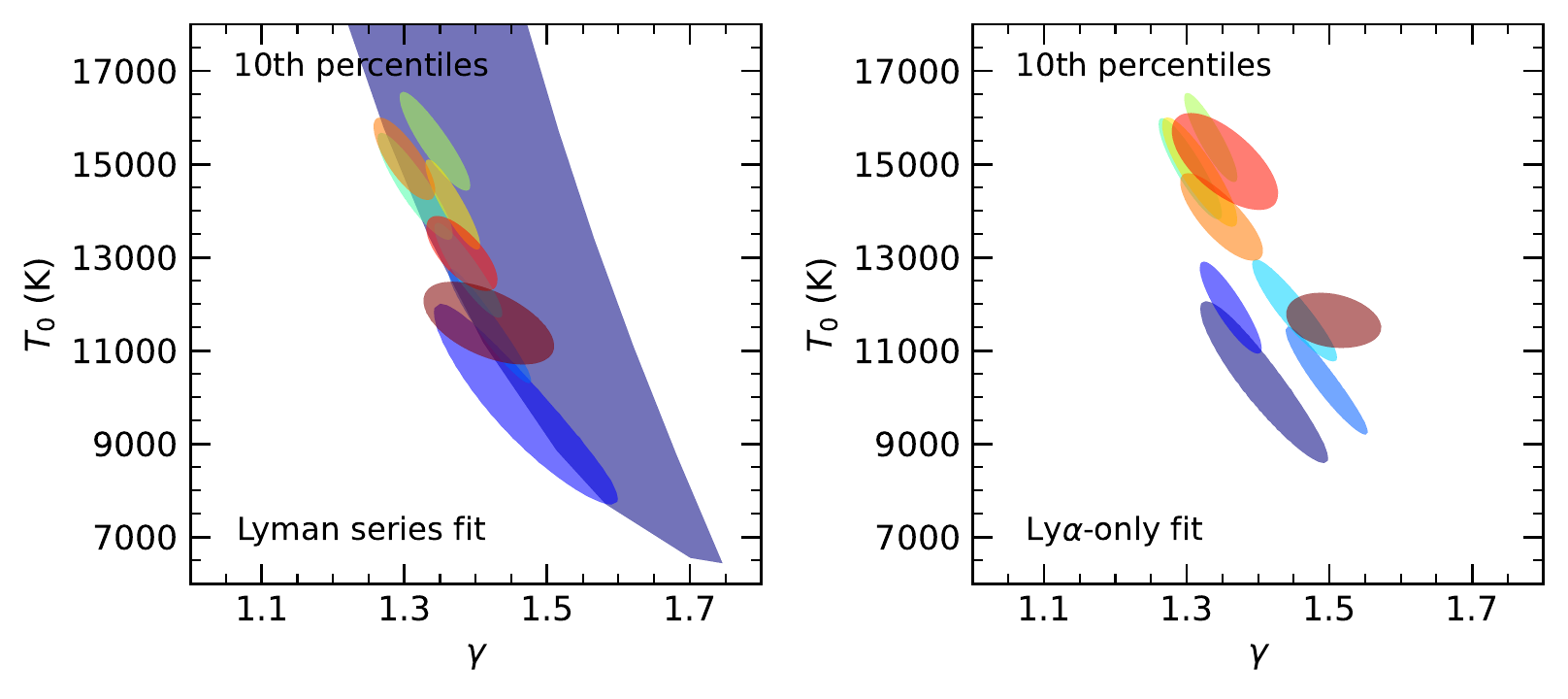}
    \includegraphics[height=7cm]{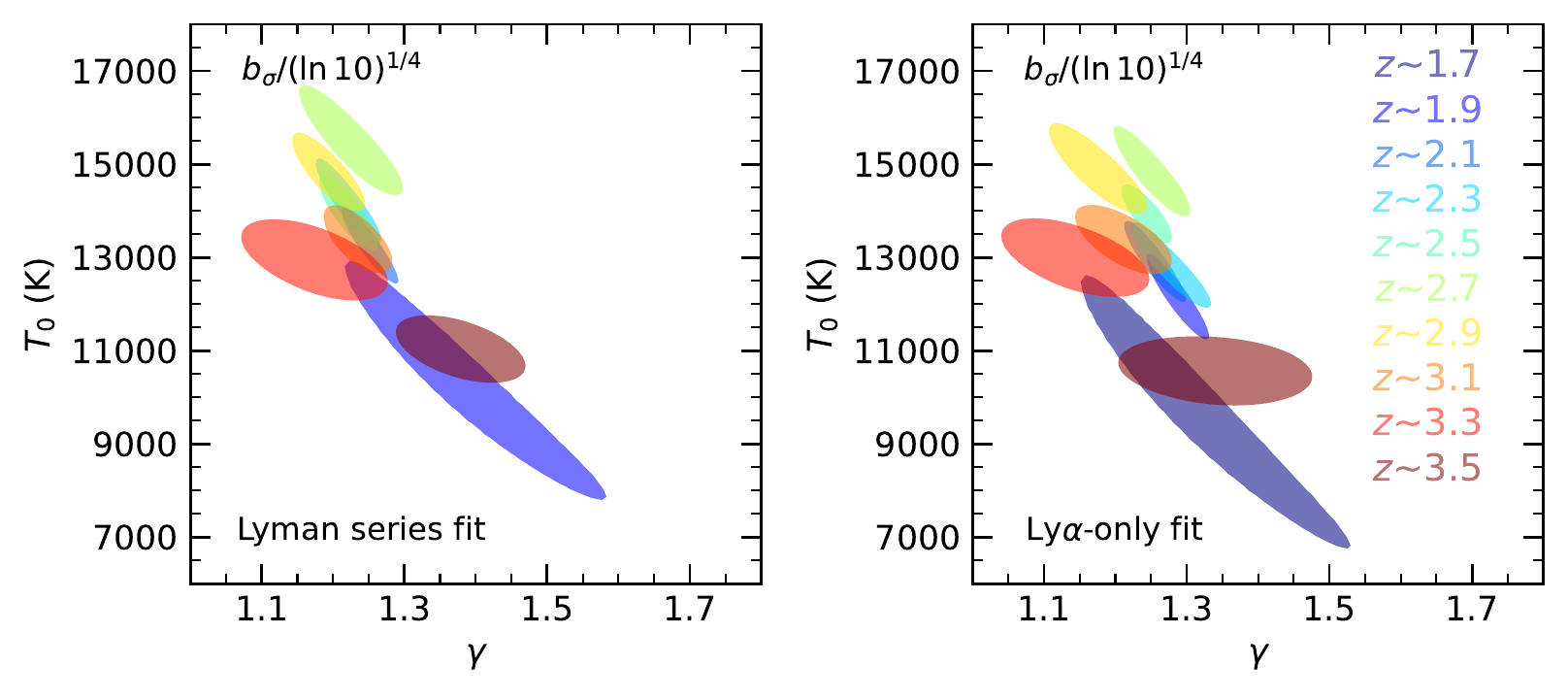}
    \caption{Joint constraints on the IGM temperature-density relation parameters, $T_0$ and $\gamma$, from $\NHI$ and $b$ values measured based on the Lyman series fit (left panels) and \lya-only fit (right panels). The constraints in the top panels are derived using the $b$ cutoff relation estimated from the data, while those in the bottom panels are from the $b$ cutoff relation inferred from fitting the $b$ distribution. The contours show the central 39\% of the distribution for the two parameters.}
    \label{fig:contour_T0gamma}
\end{figure*}

In Figure~\ref{fig:contour_T0gamma}, we show the constraints in the $T_0$--$\gamma$ plane for different redshift bins, using $b$ cutoff profile estimated non-parametrically from the data (top panels) and parametrically from fitting the $b$ distribution, with absorber $\NHI$ and $b$ measurements based on the Lyman series fit (left panels) and the \lya-only fit (right panels). As discussed in Section~\ref{sec:sensitivity}, with the relation between the $b$ cutoff and $\NHI$, the constraints on $T_0$ are mainly from the relation's amplitude, while those on $\gamma$ come from its shape. At each redshift, the constraints in the $T_0$--$\gamma$ show a degeneracy direction: a higher $T_0$ is compensated by a lower $\gamma$. This is easy to understand -- with a higher $T_0$ (hence higher amplitude from the model), a lower $\gamma$ value can tilt the model so that the amplitude of the $b$ cutoff profile towards high $\NHI$ can be lowered. 

%
%
\begin{table}
\centering
\caption{Constraints on $T_0$ and $\gamma$ at different redshifts, using $\NHI$-dependent $b$ cutoff profile estimated non-parametrically from the cumulative $b$ distribution (``10th percentiles'') and parametrically from fitting the full $b$ distribution (``$b_\sigma/(\ln 10)^{1/4}$''), with absorber $\NHI$ and $b$ measurements based on the Lyman series fit and the \lya-only fit.}
\label{table_T0gamma}
\begin{tabular}{@{}l*{5}{r}}
\hline
\noalign{\smallskip}
    &  \multicolumn{2}{c}{Lyman series fit} & \multicolumn{2}{c}{\lya-only fit} \\
$z$ & \multicolumn{1}{c}{$T_0/({\rm 10^4 K})$}  & \multicolumn{1}{c}{$\gamma$}  &
\multicolumn{1}{c}{$T_0/({\rm 10^4 K})$} & \multicolumn{1}{c}{$\gamma$} \\
\noalign{\smallskip}
\hline 
\noalign{\smallskip}
10th percentiles & & & & \\
\noalign{\smallskip}
\hline 
\noalign{\smallskip}
1.7 $\pm$ 0.1 & 2.14 $\pm$ 1.40 & 1.288 $\pm$ 0.544 & 1.03 $\pm$ 0.17 & 1.404 $\pm$ 0.088 \\ 
1.9 $\pm$ 0.1 & 0.97 $\pm$ 0.21 & 1.468 $\pm$ 0.126 & 1.19 $\pm$ 0.10 & 1.361 $\pm$ 0.043 \\ 
2.1 $\pm$ 0.1 & 1.15 $\pm$ 0.12 & 1.419 $\pm$ 0.058 & 1.03 $\pm$ 0.12 & 1.495 $\pm$ 0.057 \\ 
2.3 $\pm$ 0.1 & 1.27 $\pm$ 0.10 & 1.388 $\pm$ 0.047 & 1.18 $\pm$ 0.11 & 1.451 $\pm$ 0.059 \\ 
2.5 $\pm$ 0.1 & 1.44 $\pm$ 0.11 & 1.317 $\pm$ 0.052 & 1.49 $\pm$ 0.11 & 1.306 $\pm$ 0.044 \\ 
2.7 $\pm$ 0.1 & 1.54 $\pm$ 0.11 & 1.343 $\pm$ 0.049 & 1.55 $\pm$ 0.09 & 1.334 $\pm$ 0.037 \\ 
2.9 $\pm$ 0.1 & 1.41 $\pm$ 0.10 & 1.368 $\pm$ 0.038 & 1.48 $\pm$ 0.12 & 1.318 $\pm$ 0.052 \\ 
3.1 $\pm$ 0.1 & 1.51 $\pm$ 0.09 & 1.301 $\pm$ 0.043 & 1.38 $\pm$ 0.09 & 1.349 $\pm$ 0.057 \\ 
3.3 $\pm$ 0.1 & 1.31 $\pm$ 0.08 & 1.381 $\pm$ 0.050 & 1.50 $\pm$ 0.10 & 1.355 $\pm$ 0.074 \\ 
3.5 $\pm$ 0.1 & 1.15 $\pm$ 0.09 & 1.420 $\pm$ 0.090 & 1.16 $\pm$ 0.06 & 1.504 $\pm$ 0.065 \\ 
\noalign{\smallskip}
\hline 
\noalign{\smallskip}
$b_\sigma/(\ln 10)^{1/4}$ & & & & \\
\noalign{\smallskip}
\hline 
\noalign{\smallskip}
1.7 $\pm$ 0.1 & -               & -                 & 0.95 $\pm$ 0.28 & 1.327 $\pm$ 0.174 \\ 
1.9 $\pm$ 0.1 & 1.02 $\pm$ 0.24 & 1.391 $\pm$ 0.169 & 1.21 $\pm$ 0.09 & 1.288 $\pm$ 0.043 \\ 
2.1 $\pm$ 0.1 & 1.33 $\pm$ 0.08 & 1.250 $\pm$ 0.038 & 1.29 $\pm$ 0.09 & 1.254 $\pm$ 0.042 \\ 
2.3 $\pm$ 0.1 & 1.41 $\pm$ 0.09 & 1.223 $\pm$ 0.045 & 1.28 $\pm$ 0.09 & 1.280 $\pm$ 0.055 \\ 
2.5 $\pm$ 0.1 & 1.39 $\pm$ 0.08 & 1.223 $\pm$ 0.041 & 1.39 $\pm$ 0.06 & 1.245 $\pm$ 0.035 \\ 
2.7 $\pm$ 0.1 & 1.54 $\pm$ 0.12 & 1.231 $\pm$ 0.071 & 1.47 $\pm$ 0.10 & 1.255 $\pm$ 0.053 \\ 
2.9 $\pm$ 0.1 & 1.48 $\pm$ 0.08 & 1.195 $\pm$ 0.051 & 1.48 $\pm$ 0.10 & 1.178 $\pm$ 0.068 \\ 
3.1 $\pm$ 0.1 & 1.34 $\pm$ 0.07 & 1.236 $\pm$ 0.048 & 1.33 $\pm$ 0.07 & 1.216 $\pm$ 0.066 \\ 
3.3 $\pm$ 0.1 & 1.29 $\pm$ 0.09 & 1.183 $\pm$ 0.098 & 1.29 $\pm$ 0.08 & 1.153 $\pm$ 0.101 \\ 
3.5 $\pm$ 0.1 & 1.10 $\pm$ 0.07 & 1.383 $\pm$ 0.088 & 1.05 $\pm$ 0.07 & 1.355 $\pm$ 0.128 \\ 
\hline
\end{tabular}
\end{table}

%
%
\bibliographystyle{raa}
\bibliography{bibtex_ly}
\label{lastpage}
\end{document}